\documentclass[aps,prc,twocolumn,superscriptaddress,floatfix,amsmath,amssymb,nofootinbib,longbibliography]{revtex4-2}

\usepackage{bm}
\usepackage{dcolumn}
\usepackage{amsfonts}
\usepackage{amsthm}
\usepackage{bbold}
\usepackage{color}
\usepackage{braket}
\usepackage{graphicx}
\usepackage{multirow}
\usepackage{siunitx}
\usepackage[caption=false]{subfig}

\usepackage{orcidlink}
\usepackage{hyperref}
\hypersetup{
    colorlinks = true,
    citecolor  = blue,
    linkcolor  = blue,
    urlcolor  = blue
}

\newcommand{\ai}{\textit{ab initio}}

\newcommand{\deltago}{$\Delta$NNLO$_\text{GO}$\,(394)}

\newcommand{\nnlosat}{NNLO$_\text{sat}$}

\newcommand{\Xm}[1]{\ensuremath{#1_\text{max}}}
\newcommand{\MeV}{\ensuremath{\text{MeV}}}

\newcommand{\mel}[3]{\langle #1 | #2 | #3 \rangle}

\begin{document}

\author{F. Bonaiti~\orcidlink{0000-0002-3926-1609}\;\textsuperscript{\S}}
\email{bonaiti@frib.msu.edu}
\affiliation{Facility for Rare Isotope Beams, Michigan State University, East Lansing, Michigan 48824, USA}
\affiliation{Physics Division, Oak Ridge National Laboratory, Oak Ridge, Tennessee 37831, USA}

\author{A. Porro~\orcidlink{0000-0001-9828-546X}\;\textsuperscript{\S}}
\email{andrea.porro@tu-darmstadt.de}
\affiliation{Technische Universit\"at Darmstadt, Department of Physics, 64289 Darmstadt, Germany}
\affiliation{ExtreMe Matter Institute EMMI, GSI Helmholtzzentrum f\"ur Schwerionenforschung GmbH, 64291 Darmstadt, Germany}

\author{S. Bacca\orcidlink{0000-0002-9189-9458}}
\affiliation{Institut f\"{u}r Kernphysik and PRISMA+ Cluster of Excellence, Johannes Gutenberg-Universit\"{a}t Mainz, 55128 Mainz, Germany}
\affiliation{Helmholtz-Institut Mainz, Johannes Gutenberg-Universität Mainz, D-55099 Mainz, Germany}

\author{A. Schwenk\orcidlink{0000-0001-8027-4076}}
\affiliation{Technische Universit\"at Darmstadt, Department of Physics, 64289 Darmstadt, Germany}
\affiliation{ExtreMe Matter Institute EMMI, GSI Helmholtzzentrum f\"ur Schwerionenforschung GmbH, 64291 Darmstadt, Germany}
\affiliation{Max-Planck-Institut f\"ur Kernphysik, Saupfercheckweg 1, 69117 Heidelberg, Germany}

\author{A.~Tichai\orcidlink{0000-0002-0618-0685}}
\affiliation{Technische Universit\"at Darmstadt, Department of Physics, 64289 Darmstadt, Germany}
\affiliation{ExtreMe Matter Institute EMMI, GSI Helmholtzzentrum f\"ur Schwerionenforschung GmbH, 64291 Darmstadt, Germany}
\affiliation{Max-Planck-Institut f\"ur Kernphysik, Saupfercheckweg 1, 69117 Heidelberg, Germany}

\title{{\it Ab initio} calculations of monopole sum rules: \\ From finite nuclei to infinite nuclear matter}


\begin{abstract}
We compute moments of the isoscalar monopole response of $N=Z$ closed-shell nuclei based on chiral nucleon–nucleon plus three-nucleon interactions. We employ the random phase approximation (RPA) and two \textit{ab initio} many-body approaches, the in-medium similarity renormalization group (IMSRG) and coupled-cluster theory (CC). 
In the IMSRG framework, the moments are obtained as ground-state expectation values, whereas in the CC approach, they are evaluated through excited-state calculations.
We find good agreement between the IMSRG and CC results across all nuclei studied. RPA provides a reasonable approximation to the correlated methods if the interaction is soft. From the calculated moments, we extract average energies of the monopole response, compute finite-nucleus incompressibilities, and  estimate the incompressibility of symmetric nuclear matter by a fit to a leptodermous expansion. Our extrapolated values are lower than those obtained in nuclear matter calculations with the same interactions, but the values are consistent with phenomenological ranges.
\end{abstract}

\maketitle

\begingroup
\renewcommand\thefootnote{\S}
\footnotetext{Francesca Bonaiti and Andrea Porro contributed equally to this work and share first authorship.}
\endgroup

\section{Introduction}

The investigation of giant resonances offers deep insights into the structure and dynamical properties of atomic nuclei~\cite{Goeke:1982qm,Harakeh01a,Garg18a,Colo22a}. In particular, isoscalar giant monopole resonances (ISGMR, or GMR for short) represent a signature of collective nuclear vibrations while also establishing a direct connection between atomic nuclei and the nuclear equation of state (EOS). Moments of the monopole strength function can be related to compression properties of finite nuclei~\cite{Blaizot1980,Roca-Maza18a,Garg18a}, and, through a leptodermous expansion, to the incompressibility of infinite nuclear matter at saturation density, a key EOS property. Through the incompressibility, these resonances thus provide constraints for matter under extreme conditions in neutron stars, core-collapse supernovae, and neutron star mergers~\cite{Lattimer:2000kb,Janka:2006fh,Yasin:2018ckc,Lattimer:2021emm,Janka:2022krt,Jacobi:2023olu,Ricigliano:2024lwf}. For example, larger values of the incompressibility typically lead to stiffer EOSs.

A microscopic understanding of the response of nuclei to external probes, and of the GMR in particular, has been a longstanding challenge in nuclear structure studies.
While historically this was studied predominantly using phenomenological energy density functionals (EDFs)~\cite{Ring1980_ManyBodyBook,Bortignon98a,Harakeh01a,Bender03a,Niksic11a,Nakatsukasa16a,Garg18a,schunck2019,Colo22a}, \ai{} advances allow to target such observables from first principles using nuclear interactions (and currents) systematically derived from chiral effective field theory~\cite{epelbaum2009,machleidt2011,hammer2020,Krebs20a}.
Combined with systematically improvable many-body expansions this allows to assess theoretical uncertainties. Among these advances for nuclear responses are the Lorentz integral transform technique combined with the coupled-cluster framework (LIT-CC)~\cite{Bacca13a,Bacca14a,Miorelli16a,Acharya24a,Bonaiti2024,Marino2025openshell}, the self-consistent Green’s function approach~\cite{Raimondi18a,Raimondi18b}, and the in-medium similarity renormalization group (IMSRG)~\cite{Parzuchowski16a,Parzuchowski17a,Porro25a}. More recently, response functions have also been addressed within a time-dependent coupled-cluster framework~\cite{bonaiti2025}. 

Efforts to implement the random-phase approximation (RPA) using the same chiral interactions employed in \textit{ab initio} methods have been carried out within symmetry-conserving schemes~\cite{Papakonstantinou17a,Wu18a,Hu20a}, and later extended to deformed nuclei through the symmetry-breaking quasiparticle RPA~\cite{Beaujeault-Taudiere22a,Zaragoza24a}, highlighting the influence of static correlations. In addition, RPA and related methods constructed on correlated ground states have stressed the need for a consistent treatment of correlations in both ground and excited configurations~\cite{Carter22a,Beaujeault-Taudiere22a}. Lastly, the projected generator coordinate method (PGCM) and its perturbatively corrected extension (PGCM-PT) have been formulated to describe rotational and vibrational collective degrees of freedom within a unified framework~\cite{Frosini21a,Frosini21b,Frosini21c,Porro24a,Porro24b,Porro24c,Porro24d}.

In this work, we present LIT-CC and IMSRG calculations of the monopole strength function and its moments. On the one hand, within the LIT-CC method, the response function is explicitly constructed from a set of excited states obtained from an equation-of-motion (EOM) approach~\cite{Hagen2014Review}. On the other hand, a newly developed approach in the IMSRG framework allows to evaluate selected moments of the strength function as ground-state expectation values~\cite{Porro25a}.

In the first part of the paper, we focus on moments of the monopole response and compare the results of the two approaches. Both the CC and IMSRG frameworks have been shown to accurately predict ground- and excited-state properties of nuclei at comparable accuracy, see, e.g.,~\cite{Parzuchowski17a,Miyagi20a,Heinz:2024juw}. In this work we demonstrate that this consistency extends to collective excitations as well. IMSRG and CC results are also compared to RPA calculations.

In the second part of the paper, we analyze the connection between properties of the monopole response and the incompressibility of nuclear matter. In particular, we calculate the finite-nucleus incompressibility, and use a leptodermous expansion to extract the nuclear matter value. The extrapolated values are lower than those obtained in nuclear matter calculations with the same interactions, but the values are consistent with phenomenological ranges.

This paper is organized as follows. In Secs.~\ref{sec:methods} and~\ref{sec:methods1}, the many-body frameworks used to calculate the nuclear response and its moments are introduced. Results are presented for selected closed-shell nuclei in Sec.~\ref{sec:finitenuclei}. 
In Sec.~\ref{sec:nuclearmatter}, the extrapolated results for the incompressibility are compared to nuclear matter calculations.
Finally, we summarize our results and give an outlook in Sec.~\ref{sec:outlook}. Further numerical benchmarks are given in the Appendix.

\section{The nuclear response}
\label{sec:methods}

\subsection{Strength functions and their moments}
\label{sec:mom-strategies}

In the following the nuclear Hamiltonian
\begin{equation}
\label{eq:Hamiltonian}
    H=T-T_{\text{CM}}+V_{\text{NN}}+V_{\text{3N}} \,,
\end{equation}
is employed, where $T$ is the total kinetic energy, $T_{\text{CM}}$ the center-of-mass (CM) kinetic energy, $V_{\text{NN}}$ denotes nucleon-nucleon (NN) and $V_{\text{3N}}$ three-nucleon (3N) interactions. The set of the eigenstates of the Hamiltonian $H$ is denoted by $\{\ket{\Psi_\nu}\}$, each of which satisfies the time-independent Schr\"odinger equation
\begin{equation}
    H\ket{\Psi_\nu} = E_\nu \ket{\Psi_\nu} \,.
\end{equation}
Given a Hermitian operator $Q$, the associated strength function (or response function) reads as
\begin{equation}
\label{eq:strength}
    S(Q,E)\equiv\sum_{\nu}|\langle\Psi_\nu|Q|\Psi_0\rangle|^2\delta(E_\nu-E_0-E)\,,
\end{equation}
which fully characterizes the linear response of the nucleus to an external perturbation $Q$.  

The moments of the response function are defined as energy-weighted averages of the transition amplitudes entering Eq.~\eqref{eq:strength} according to
\begin{align}
\label{eq:mom_sum}
    m_k(Q)&\equiv\int_0^\infty E^kS(Q,E) \, dE \,, \nonumber\\
    &=\sum_{\nu}(E_\nu-E_0)^k|\langle\Psi_\nu|Q|\Psi_0\rangle|^2\,.
\end{align}
If $Q$ has a non-vanishing expectation value in the ground state, the elastic contribution to the total strength is subtracted by replacing $Q$ with $Q - \braket{\Psi_0|Q|\Psi_0}$.
Different moments may be used to compactly characterize the response of the system. In particular, several average energies can be introduced as
\begin{subequations}
\label{eq:aves}
    \begin{align}
    \tilde{E}_k(Q)&\equiv\sqrt{\frac{m_k(Q)}{m_{k-2}(Q)}}\,,\\
    \bar{E}_k(Q)&\equiv\frac{m_k(Q)}{m_{k-1}(Q)}\,.
    \end{align}
\end{subequations}
These average energies obey a series of inequalities
\begin{equation}
\label{eq:ineq}
    \ldots\leq\tilde{E}_k(Q)\leq\bar{E}_k(Q)\leq\tilde{E}_{k+1}(Q)\leq\bar{E}_{k+1}(Q)\leq\ldots\,.
\end{equation}
If the entire strength were concentrated in a single peak, the equality sign in Eq.~\eqref{eq:ineq} would hold. The spread of the different averages in Eqs.~\eqref{eq:aves} is a quantitative measure of the fragmentation of the strength $S(Q,E)$.
Note that the evaluation of Eq.~\eqref{eq:mom_sum} formally requires the knowledge of the entire spectrum of $H$, i.e., one needs access to all the bound and continuum excited states of the system $\ket{\Psi_\nu}$. In this work, two strategies that are theoretically equivalent, though potentially different in practical implementation, are employed to calculate the moments, as explained below. 

\subsection{Moments as ground-state expectation values}
\label{sec:gs_mom}

Equation~\eqref{eq:mom_sum} can be rewritten in terms of the expectation value on the ground state $\ket{\Psi_0}$,
\begin{equation}
    m_k(Q)=\braket{\Psi_0|M_k(Q)|\Psi_0}\,,
\end{equation}
where the so-called moment operator is
\begin{equation}
    M_k(Q)\equiv Q^\dagger[\underbrace{{H},...[{H},[{H}}_{\text{$k$ times}},{Q}]]...]\,.
\end{equation}
The latter is defined through a $k$-fold nested commutator involving the full nuclear Hamiltonian~\cite{Bohigas79a}.
Here, we used the relation
\begin{equation}
    (H-E_0)^k\ket{\Psi_0}=0\,,
\end{equation}
together with the completeness relation
\begin{equation}
\label{eq:complete}
    \sum_{\nu} \ket{\Psi_{\nu}} \bra{\Psi_{\nu}} = 1\,.
\end{equation}
By further exploiting the Hermiticity of the operator, odd moments  can be written as
\begin{equation}
        M_k(Q)=\frac{1}{2} [Q,[\underbrace{{H},...[{H},[{H}}_{\text{$k$ times}},{Q}]]...]]]\,.
\end{equation}
The explicit expression for the lowest-$k$ moments are given by
\begin{subequations}
\label{eq:moms_op}
    \begin{align}
        M_0(Q)&\equiv Q^2\,,\label{eq:m0_op}\\
        M_1(Q)&=\frac{1}{2}[Q,[H,Q]]\label{eq:double_comm}\,.
    \end{align}
\end{subequations}
The former discussion is, in principle, valid for any integer value of $k$, either positive or negative. However, negative values of $k$ require the definition of an inverse operation for the commutator, which in turn requires the knowledge of the exact spectrum of $H$. However, negative moments can also be evaluated in terms of polarizabilities, which can be easily shown employing first-order perturbation theory (dielectric theorem)~\cite{Bohigas79a,Capelli09a}. Eventually, the following formulas are found
\begin{subequations}
\label{eq:pols}
    \begin{align}
    \label{eq:pol_Q}
        m_{-1}(Q)&=-\frac{1}{2}\frac{\partial}{\partial\lambda}\mathcal{Q}_\lambda\Big|_{\lambda=0}\,,\\
        m_{-1}(Q)&=\frac{1}{2}\frac{\partial^2}{\partial\lambda^2}\mathcal{H}_\lambda\Big|_{\lambda=0}\,,
    \end{align}
\end{subequations}
where
\begin{subequations}
    \begin{align}
        \mathcal{Q}_\lambda&\equiv\braket{\Psi(\lambda)|Q|\Psi(\lambda)}\,,\\
        \mathcal{H}_\lambda&\equiv\braket{\Psi(\lambda)|H|\Psi(\lambda)}\,,
    \end{align}
\end{subequations}
and $\ket{\Psi(\lambda)}$ is the solution to the perturbed Schr\"odinger equation
\begin{equation}
    (H+\lambda\,Q)\ket{\Psi(\lambda)}\equiv H_{\lambda}\ket{\Psi(\lambda)}=E(\lambda)\ket{\Psi(\lambda)}\,,
\end{equation}
with $\lambda$ the perturbation parameter. As displayed by Eqs.~\eqref{eq:pols}, a perturbation of order $\lambda$ produces first-order changes in the expectation value of the perturbation operator $Q$ itself, while second-order changes $\mathcal{O}(\lambda^2)$ are obtained in the energy. Hence, in the evaluation of derivatives entering Eqs.~\eqref{eq:pols}, Eq.~\eqref{eq:pol_Q} is considered to be numerically more robust.

\subsection{Moments from integral transforms}
\label{sec:momentslit}

To avoid the explicit calculation of bound and continuum excited states of the Hamiltonian, one can choose to adopt the LIT method that is based on an integral transform of the strength function with Lorentzian kernel~\cite{efros2007},
\begin{equation}
L(\sigma, \Gamma) = \frac{\Gamma}{\pi} \int dE\;\frac{S(Q, E)}{(E - \sigma)^2 + \Gamma^2} \,,
\label{eq:lit}
\end{equation}
where $\sigma$ is the centroid and $\Gamma$ is the width of the kernel. By inserting Eq.~\eqref{eq:strength} into Eq.~\eqref{eq:lit} and using the completeness relation from Eq.~\eqref{eq:complete} one obtains
\begin{equation}
\begin{split}
    L(\sigma, \Gamma) = &\frac{\Gamma}{\pi} \bra{\Psi_0} Q\frac{1}{H - E_0 - \sigma + i\Gamma} \\ &\times\frac{1}{H - E_0 - \sigma - i\Gamma} Q\ket{\Psi_0} \,. 
\end{split}
\end{equation}
Defining the complex variable $z \equiv E_0+ \sigma + i\Gamma$, the integral transform can then be computed as 
\begin{equation}
    L(z) = \frac{\Gamma}{\pi} \braket{\Tilde{\Psi}|\Tilde{\Psi}} \,,
\end{equation}
where $\ket{\Tilde{\Psi}}$ represents the solution to the equation
\begin{equation}
    (H-z)\ket{\Tilde{\Psi}} = Q \ket{\Psi_0}
    \label{eq:boundstatelike}
\end{equation}
for different values of $\sigma$ and $\Gamma$. In this way, the task of computing the full spectrum of the Hamiltonian is reduced to solving a Schr\"odinger-like equation~\eqref{eq:boundstatelike}, for which bound-state approaches can be applied. Ultimately, the moments of the strength function are evaluated in the limit $\Gamma \to 0$, where the Lorentzian kernel approaches a Dirac $\delta$-distribution. In this limit, the LIT reduces to the strength function
\begin{equation}
    L(\sigma, \Gamma\rightarrow 0) = \int dE\;S(Q, E) \delta(E - \sigma) = S(Q, \sigma) \,.
\end{equation}
Such a strength function is discretized, in the sense that the information related to the excited states in the continuum is represented by bound pseudo-states, and it allows for the evaluation of moments via 
\begin{equation}
    m_k(Q) = \int d\sigma\; \sigma^k L(\sigma, \Gamma\rightarrow 0) \,.
    \label{eq: moments}
\end{equation}

\section{Many-body frameworks}
\label{sec:methods1}

\subsection{Moment operators from the IMSRG}

In the IMSRG framework~\cite{Tsukiyama10a,Hergert15a} the initial Hamiltonian is evolved via a continuous unitary transformation $U(s)$ parametrized by a flow parameter $s$.
This aims for the decoupling of $n$-particle-$n$-hole ($n$p-$n$h) excitations from a many-body reference state,
\begin{equation}
    H(s)=U(s)H(0)U^\dagger(s) \,.
\end{equation}
The transformation can be reformulated as an ordinary differential equation involving the evaluation of a commutator between two many-body operators,
\begin{equation}
    \frac{dH(s)}{ds}=[\eta(s),H(s)]\,,
\end{equation}
where $\eta(s)$ is the (anti-Hermitian) generator of the transformation
\begin{equation}
    \eta(s) \equiv \frac{d\,U(s)}{ds}U^\dagger(s)=-\eta^\dagger(s)\,.
\end{equation}
In this work the generator is chosen as the imaginary-time generator, which is widely used in applications to closed-shell nuclei.

The initial many-body reference state $\ket{\Phi_0}$ is taken to be a symmetry-conserving Hartree-Fock (HF) Slater determinant. All operators are then normal ordered with respect to $\ket{\Phi_0}$, such that the zero-body component of the Hamiltonian after normal ordering is given by its expectation value $\braket{\Phi_0|H(0)|\Phi_0}$. Through the solution of the IMSRG flow the coupling of $\ket{\Phi_0}$ to $n$p-$n$h excitations is suppressed, and the zero-body term approaches the exact ground-state energy in the large $s$ limit when the IMSRG equations are not truncated:
\begin{equation}
    \lim_{s\to\infty} \braket{\Phi_0|H(s)|\Phi_0}=E_0\,.
\end{equation}
In this work, the Magnus formulation of the unitary transformation is used~\cite{Morris15a}, where the latter is directly expressed by an exponential ansatz
\begin{equation}
    U(s)=e^{\Omega(s)} \,.
\end{equation}
This has the advantage of manifestly enforcing unitarity even in presence of truncations. All commutators are truncated at the normal-ordered two-body level giving rise to the IMSRG(2) truncation. While three-body operators can be included at a substantial computational cost, the IMSRG(2) approximation has been shown to yield reliable results for ground-state properties of medium-mass nuclei~\cite{Heinz:2024juw}.

With this, the transformed Hamiltonian, as well as any other operator $O$, can be constructed as
\begin{subequations}
    \begin{align}
        H(s)&=e^{\Omega(s)}He^{-\Omega(s)}\,,\\
        O(s)&=e^{\Omega(s)}Oe^{-\Omega(s)}\,.
    \end{align}
\end{subequations}
This also extends to the moment operators defined in Eqs.~\eqref{eq:moms_op}, such that the moments expectation value at the HF and IMSRG level are given by
\begin{subequations}
    \label{eq:used_mom_def}
    \begin{align}
        m_k(Q_\lambda,0)&=\braket{\Phi_0|M_k(Q_\lambda)|\Phi_0}\,,\\
        m_k(Q_\lambda,s)&=\braket{\Phi_0|e^{\Omega(s)}M_k(Q_\lambda)e^{-\Omega(s)}|\Phi_0}\,,
    \end{align}    
\end{subequations}
respectively. Thus, differences in the value of $m_k$ before (HF) and after (IMSRG) the flow reflect the impact of ground-state correlations on the nuclear response.

\subsection{Integral transforms in coupled-cluster theory}

In CC theory~\cite{Hagen2014Review,ShavittBartlett} one starts from an exponential ansatz for the many-body ground state,
\begin{align}
    \label{eq: cc gs ansatz}
    \ket{\Psi_0} = e^{T} \ket{ \Phi_0 } \,.
\end{align}
Correlations are introduced through 
$e^{T}$, where the cluster operator $T$ is expanded in a truncated $n$p-$n$h operator basis.
In the singles and doubles level (CCSD) approximation, the cluster operator consists of 
\begin{equation}
    T= \sum_{ai} t^{a}_{i} a_a^{\dagger}a_i +\frac{1}{4} \sum_{abij} t^{ab}_{ij} a_a^{\dagger} a_b^{\dagger} a_j a_i \,,
\end{equation}
where $a^\dagger$ and $a$ are creation and annihilation operators, while the indices $i,j$ and $a,b$ denote hole and particle states, respectively.
After normal ordering the Hamiltonian with respect to the reference state, one introduces the similarity-transformed Hamiltonian
\begin{align}
    \overline{H} \equiv e^{-{T}} H e^{{T}} \,,
\end{align}
and computes the amplitudes $t^{a}_{i}, t^{ab}_{ij} $ by solving a set of non-linear amplitude equations
\begin{align}
\nonumber
    \mel{ \Phi_{i}^{a} }{ \overline{H} }{ \Phi_0 } & = 0, \\
    \mel{ \Phi_{ij}^{ab} }{ \overline{H} }{ \Phi_0 } &= 0\,,
\end{align}
which decouple particle-hole excitations from the reference state $|\Phi_0\rangle$.
Additional higher-body correlations can be approximately included to achieve greater precision, using, e.g., the CCSDT-1 approach~\cite{watts1993}, which accounts for the leading effects of three-particle--three-hole excitations in the nuclear ground state.

In the CC framework, excited states can be described through EOM techniques~\cite{ShavittBartlett,Hagen2014Review}.
Since similarity transformed operators in CC theory are non-Hermitian~\cite{ShavittBartlett}, both the right and left ground states are required.
Assuming that the general eigenstates $\ket{ \Psi_\nu }$ of the Hamiltonian can be constructed by acting linearly on the CC ground state with an excitation operator, the right and left eigenstates can be written as
\begin{subequations}
\begin{align}
    \label{eq: eom right ansatz}
    \ket{ \Psi_{\nu} } = { R }_{\nu}  \ket{\Psi_0} = { R }_{\nu} e^{ {T} } \ket{ \Phi_0 } \,,
    \\ 
    \label{eq: eom left ansatz}
    \bra{ \Psi_{\nu} } = \bra{ \Psi_0 } {L}_{\nu} = \bra{ \Phi_0 }  {L}_{\nu}e^{ - {T} } \,.
\end{align}%
\label{eq:eom}%
\end{subequations}
Truncating ${ R }_\nu$ and ${ L }_\nu $ at the 2p-2h level yields the following operators
\begin{subequations}
\begin{align}
    { R }_\nu &= r_0 + \sum_{ai} r^{a}_{i} a^{\dagger}_{a} a_i + \frac{1}{4} \sum_{abij} r^{ab}_{ij} a^{\dagger}_{a} a^{\dagger}_{b} a_{j} a_{i} \,, 
    \\
    { L }_\nu &= l_0 + \sum_{ai} l^{i}_{a} a^{\dagger}_{i} a_{a} + \frac{1}{4} \sum_{abij} l_{ab}^{ij} a^{\dagger}_{i} a^{\dagger}_{j} a_{b} a_{a} \,.
\end{align}
\end{subequations}
In the special case of the nuclear ground state $\nu=0$, the amplitudes reduce to ${R}_0 = 1$ and ${L}_{0} = 1 + {\Lambda}$.
Using the EOM ansatz [Eqs.~\eqref{eq:eom}], we can finally rewrite the strength function in the CC formalism as
\begin{equation}
\begin{split}
   S(Q, E) &= \sum_{\nu} \langle \Phi_0 \vert L_0 \overline{Q}^\dagger R_\nu \vert \Phi_0 \rangle \langle \Phi_0 \vert L_\nu \overline{Q} \vert \Phi_0 \rangle \\
   &\quad \times \delta(E_{\nu} - E_0 - E) \,,
   \label{eq: cc strength}
\end{split}
\end{equation}
where the similarity-transformed transition operator 
\begin{equation}
    \overline{Q} = e^{-{T}} Q e^{{T}}
\end{equation}
was introduced. Starting from Eq.~(\ref{eq: cc strength}), one can evaluate moments by coupling CC theory with the LIT technique of Sec.~\ref{sec:momentslit} in the so-called LIT-CC method~\cite{Bacca13a,Bacca14a}. In this approach, Eq.~\eqref{eq:boundstatelike} becomes an EOM-CC problem with a source term.

\begin{figure*}[t]
    \centering
    \includegraphics[width=\linewidth]{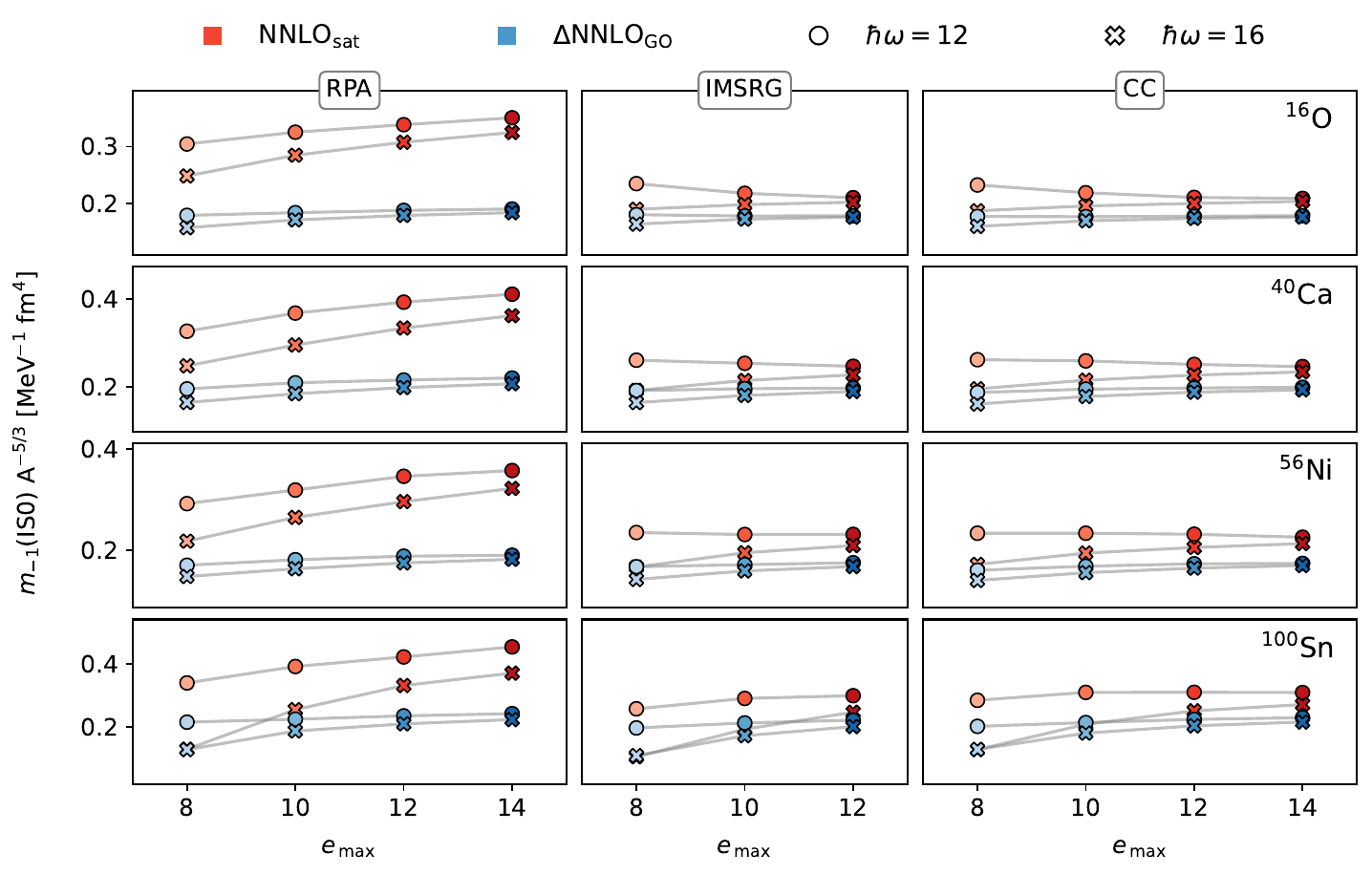}
    \caption{Values of $m_{-1}$ for the isoscalar monopole (IS0) response for RPA (left), IMSRG (middle), and CC (right) calculations. Results are given for two different interactions (\nnlosat{}~\cite{Ekstrom15a}, \deltago{}~\cite{Jiang20a}) for $\hbar\omega=12-16$\,MeV as a function of the model space size \Xm{e}. Absolute values are rescaled by a factor $A^{5/3}$ in order to show all nuclei on the same scale.}
    \label{fig:mm1}
\end{figure*}

\section{Applications to finite nuclei}
\label{sec:finitenuclei}

In this work, we focus on the isoscalar monopole (IS0) response, whose excitation operator reads
\begin{align}
\label{eq:operator}
    Q(\text{IS}0)&=\sum_{i=1}^Ar_i^2 \,,
\end{align}
where $r_i$ is the radial component of the position of the $i$-th particle in the laboratory frame.

\subsection{Computational setup}
\label{sec:setup}

In this work, we study the nuclear response for two different chiral NN and 3N interactions: the \nnlosat{} interaction from Ref.~\cite{Ekstrom15a} and the \deltago{} interaction from Ref.~\cite{Jiang20a}. All operators are expanded in spherical harmonic-oscillator (HO) basis states consisting of at most 15 major shells, i.e., $\Xm{e} = \text{max}(2n + l) = 14$. Three-nucleon interactions are included using the normal-ordered two-body approximation. This has been shown to introduce only small errors in medium-mass nuclei~\cite{Hagen2007,Roth12,Heinz:2024juw}.
The number of three-body configurations is additionally restricted to $e_1 + e_2 + e _3 \leq E_\text{3,max}=24$ which has been shown to be sufficient to ensure convergence in heavy systems~\cite{Miyagi2022PRC_NO2B}. The underlying HO frequency is taken in a range $\hbar \omega = 12-16$\,MeV. The NN and 3N matrix elements are computed using the \textsc{NuHamil} code~\cite{Miyagi2023EPJA_NuHamil}.  

In the following, the moments of the isoscalar monopole strength for $k=-1,0,1$ are evaluated for $N=Z$ doubly closed-shell nuclei $^{16}$O, $^{40}$Ca, $^{56}$Ni, and $^{100}$Sn. The IMSRG results are obtained by employing the ground-state-based technique described in Sec.~\ref{sec:gs_mom} in the IMSRG(2) approximation. CC calculations, instead, exploit the excited-states formulation, as discussed in Sec.~\ref{sec:momentslit}. In the latter framework, many-body truncations are performed at both the ground- and excited-state level. Unless otherwise specified, in both cases we adopt the CCSD approximation. For comparison, results from RPA calculations are also presented.

\begin{figure*}[t]
    \centering
    \includegraphics[width=\linewidth]{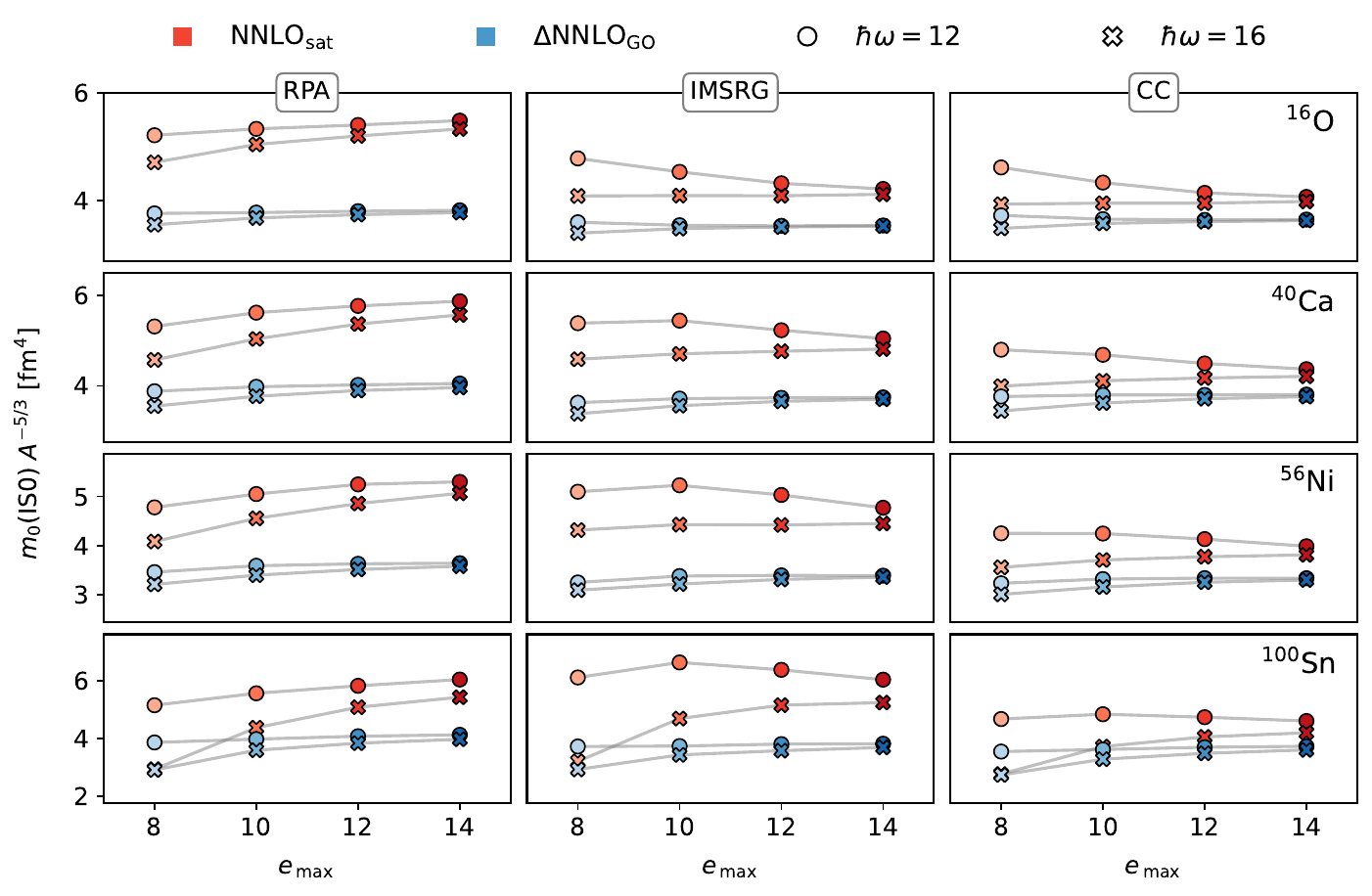}
    \caption{Values of $m_0$ for the isoscalar monopole (IS0) response for RPA (left), IMSRG (middle), and CC (right) calculations. Results are given for two different interactions (\nnlosat{}~\cite{Ekstrom15a}, \deltago{}~\cite{Jiang20a}) for $\hbar\omega=12-16$\,MeV as a function of the model space size \Xm{e}. Absolute values are rescaled by a factor $A^{5/3}$ in order to show all nuclei on the same scale.}
    \label{fig:m0}
\end{figure*}

\begin{figure*}[t]
    \centering
    \includegraphics[width=\linewidth]{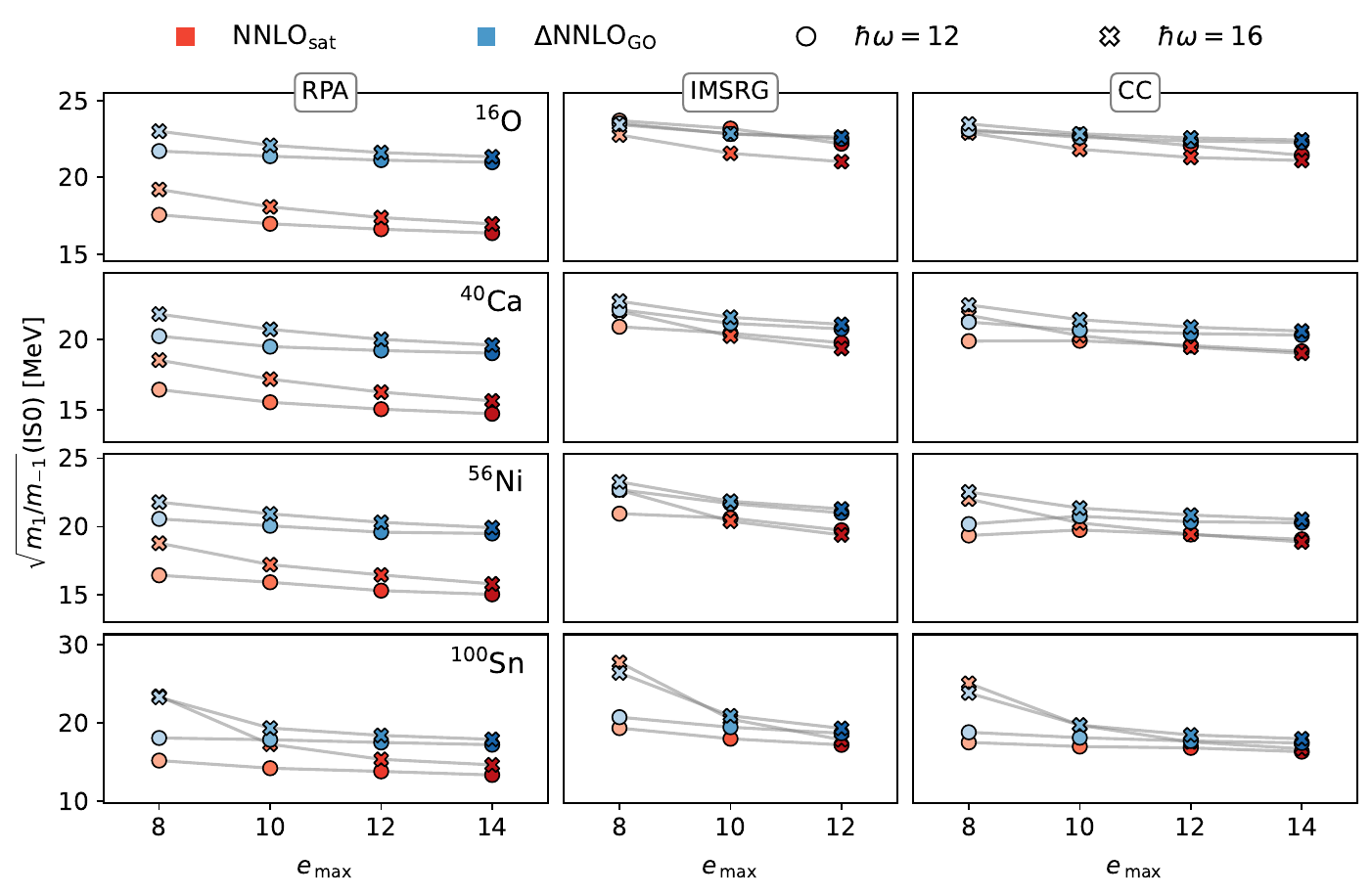}
    \caption{Average energy $\sqrt{m_1/m_{-1}}$ for the isoscalar monopole (IS0) response for RPA (left), IMSRG (centre) and CC (right) calculations. Results are given for two different interactions (\nnlosat{}~\cite{Ekstrom15a}, \deltago{}~\cite{Jiang20a}) for the HO basis frequencies $\hbar\omega=12-16$\,MeV as a function of the model space size \Xm{e}.}
    \label{fig:ave}
\end{figure*}

\subsection{Isoscalar monopole moments convergence}
\label{sec:moms_num}

We start by presenting the results relative to the moments of the IS0 operator for the different many-body methods and Hamiltonians as a function of the HO basis size and frequency. In Fig.~\ref{fig:mm1} the values of $m_{-1}$ are displayed. One can easily observe a striking difference between the two interactions. Results obtained with \nnlosat{} display a slower convergence with respect to \Xm{e}, a stronger dependence on the selected $\hbar\omega$, and on the many-body method employed. RPA results, for instance, do not display a clear sign of convergence. IMSRG and CC results, which quantitatively agree with each other, show a better \Xm{e} convergence, despite a significant $\hbar\omega$ dependence for the heaviest systems considered: relative differences amount to 13\% and 19\% in CC and IMSRG calculations, respectively. The situation is drastically different for the \deltago{} results. IMSRG, CC, and RPA results display relative deviations of only 3\%. This shows that correlations encoded in IMSRG and CC results are effectively cancelled for this observable with soft interactions, as suggested in previous works~\cite{Beaujeault-Taudiere22a}. In this specific situation, RPA can serve as a sufficiently good first approximation.
Moreover, we observe that the model-space convergence is significantly improved for the soft interaction \deltago{}, with a relative difference ranging from 1\% to 3\% in \Xm{e} and of the order of 5\% for $\hbar\omega$ variations even in heavier systems.

A similar situation is found when discussing the results for $m_0$, which are shown in Fig.~\ref{fig:m0}. When employing the \nnlosat{} interaction both the many-body and the model-space uncertainty are significantly larger. It is interesting to remark that in this case one also observes a non-negligible difference between the CC and IMSRG results, which increases for heavier nuclei. This difference is of about 20\% in $^{100}$Sn (while only of 3\% in $^{16}$O). One possible explanation relates to the more difficult convergence of even moments than odd ones, also observed in Refs.~\cite{Rowe2010a,Porro25a}. Moreover, we emphasize that in this case $m_0$ describes the variance of the mean-square radius operator. The latter depends on the quality of the ground state of the system, and as a consequence on the many-body convergence of the solution, which can be strongly affected by the use of a harder interaction such as \nnlosat{}. To further elucidate this, we considered the inclusion of triples correction in the ground state for CC calculations. Due to computational cost, we limited this analysis to $^{56}$Ni with \Xm{e} = 10 and $\hbar\omega = 16$\,MeV. Even though this \Xm{e} value is too small to reach convergence, we observe that the contribution of 3p-3h excitations improves the comparison with IMSRG(2) results. The $m_0$ moment increases from $3040$\,fm$^4$ to $3160$\,fm$^4$, reducing the difference with the IMSRG(2) result of $3630$\,fm$^4$ from 17\% to 13\% in $^{56}$Ni. IMSRG(3) corrections have been found to be small for this quantity~\cite{heinz_priv}. As already observed for $m_{-1}$, the use of the \deltago{} interaction largely reduces the uncertainty of $m_0$ associated to the model-space convergence, of the order of 1\% or smaller for \Xm{e} and from 1\% to 3\% for $\hbar\omega$ variations, and the spread between the three different methods, lying always below 5\%. 

The convergence pattern of $m_1$ does not differ significantly from the previously discussed cases, with much better convergence achieved for the softer \deltago{} interaction than with \nnlosat{}. This is shown in the Appendix for completeness.

\subsection{Average energies}
\label{sec:centroids}

As discussed in Sec.~\ref{sec:mom-strategies}, ratios of the isoscalar monopole moments can be used to estimate the average energy of the GMR $E_{\rm GMR}$. Since this will enter the finite-nuclei incompressibility, we define $E_{\rm GMR}$ as
\begin{equation}
    E_{\rm GMR} = \sqrt{\frac{m_1}{m_{-1}}} \,.
    \label{eq: egmr}
\end{equation}
Theoretical predictions for $E_{\rm GMR}$ are displayed in Fig.~\ref{fig:ave}. A good agreement between the IMSRG and CC calculations is observed across all nuclei. The \deltago{} interaction appears to favor larger values for $E_{\rm GMR}$ with respect to \nnlosat{}. Results from \deltago{} lie between 7\% and 9\% (5\% and 7\%) higher than the \nnlosat{} values for IMSRG (CC) calculations. Also, we notice that average energies exhibit faster convergence compared to the individual moments, even in the case of the harder \nnlosat{} interaction. Since $E_{\rm GMR}$ is obtained as a ratio, systematic uncertainties cancel out in the resulting values. Differences between the two interactions are less pronounced in IMSRG and CC than in RPA. The latter displays the same behavior observed for the individual moments: while it provides a sufficiently good approximation to the correlated IMSRG and CC results for the softer \deltago{} interaction, it systematically underpredicts them for \nnlosat{}.

\begin{table*}[t]
\caption{IMSRG and CC predictions for the average energies of $^{16}$O, $^{40}$Ca, $^{56}$Ni, and $^{100}$Sn with the \nnlosat{} and \deltago{} interactions, in comparison to the experimental data of Ref.~\cite{lui2001,youngblood2007,monrozeau2008}. Note that experimental data are only available in limited excitation energy ranges, see text for details. All results are given in MeV.}
\label{tab:1}
\begin{ruledtabular}
\begin{tabular}{rccccc}
\multirow{2}{*}{Nucleus} &
\multicolumn{2}{c}{\nnlosat{}} &
\multicolumn{2}{c}{\deltago{}} & \\
& {IMSRG} & {CC} & {IMSRG} & {CC} & {Experiment} \\
\hline
$^{16}$O & 21.6(8) & 21.3(4) & 22.6(1) & 22.4(1) & 19.63(38)~\cite{lui2001} \\
$^{40}$Ca & 19.5(4) & 19.1(2) & 20.9(3) & 20.4(2) & 17.29(12)~\cite{youngblood2007} \\
$^{56}$Ni & 19.6(5) & 19.0(2) & 21.2(3) & 20.4(2) & 19.3(5)~\cite{monrozeau2008} \\
$^{100}$Sn & 17.5(13) & 16.5(5) & 19.0(8) & 17.7(4) & \multicolumn{1}{c}{--} \\
\end{tabular}
\end{ruledtabular}
\end{table*}

For $^{16}$O, $^{40}$Ca, and $^{56}$Ni, experimental data for the monopole response and average energies are available, allowing for a comparison with the theoretical predictions, see Table~\ref{tab:1}. In Table~\ref{tab:1} we list the experimental average energies, defined according to Eq.~(\ref{eq: egmr}), taken from Ref.~\cite{lui2001} for $^{16}$O and from Ref.~\cite{youngblood2007} for $^{40}$Ca. For $^{56}$Ni, we include the experimental value from Ref.~\cite{monrozeau2008}, which however corresponds to the centroid energy $m_1/m_0$. The IMSRG and CC results presented in Table~\ref{tab:1} are obtained as follows. The central values are the average of the results for the two different HO frequencies $\hbar\omega$ at the largest model-space truncations $e_{\rm max}$ used for each method. The quoted uncertainty is the standard deviation among four values, corresponding to the two largest $e_{\rm max}$ truncations and the two different $\hbar\omega$, thereby reflecting the residual dependence on the model-space parameters. We emphasize that this uncertainty does not include the Hamiltonian or EFT truncation uncertainty and also not the many-body uncertainty.

Overall, the theoretical predictions for $E_{\rm GMR}$ are systematically larger than the experimental values for both many-body methods and interactions, with \nnlosat{} being closer to experiment. In particular, the \nnlosat{} results overlap with the experimental centroid energy $m_1/m_0$ for $^{56}$Ni. It is worth noting that the experimental determination of $E_{\rm GMR}$ is based on monopole strength distributions measured over finite excitation-energy ranges: $11$–$40$\,MeV for $^{16}$O~\cite{lui2001}, $7.5$–$28.8$\,MeV for $^{40}$Ca~\cite{youngblood2007}, and $0$–$40$\,MeV for $^{56}$Ni~\cite{monrozeau2008}. Since $m_{-1}$ ($m_1$) is most sensitive to the low- \mbox{(high-)} energy part of the spectrum, truncating the integration range at low (high) energies can lead to an underestimation of the average energy.
This assumption can be tested within the CC framework, where the integration limits in Eq.~\eqref{eq: moments} can be adjusted to match the experimental intervals. Focusing on $^{16}$O and $^{40}$Ca with the \nnlosat{} interaction, the average energy values are $19.93(15)$\,MeV and $18.3(3)$\,MeV, respectively. In both cases, the theoretical values move closer to experiment, with the result for $^{16}$O becoming fully consistent with data.

\section{Infinite matter incompressibility}
\label{sec:nuclearmatter}

\begin{figure*}[t]
    \centering
    \includegraphics[width=\linewidth]{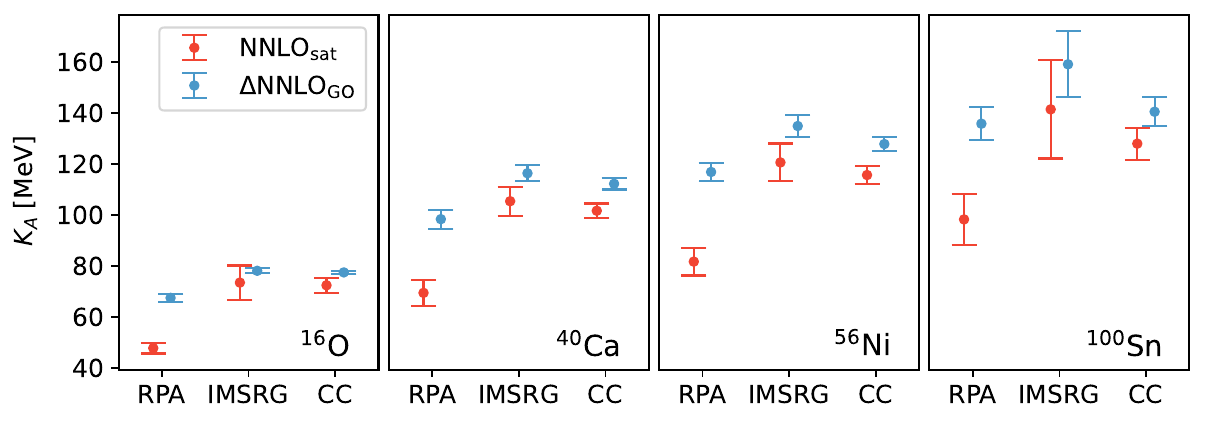}
    \caption{Finite-nucleus incompressibility $K_A$ from Eq.~\eqref{eq:KA}. Calculations were performed employing the \nnlosat{}~\cite{Ekstrom15a} and \deltago{}~\cite{Jiang20a} interactions. Numerical values are given by the average between calculations employing different $\hbar\omega$ at \Xm{e} = 14 for RPA and CC and \Xm{e} = 12 for IMSRG calculations. The error bars reflect the model-space convergence of the results and are determined from the $\hbar\omega$ variation and the two largest employed \Xm{e}.}
    \label{fig:KA_method}
\end{figure*}

Extensive studies exist linking giant monopole resonances to the incompressibility of symmetric nuclear matter, see Refs.~\cite{Blaizot1980,Jennings1980,Garg18a,Roca-Maza18a} for reviews. In infinite matter, the incompressibility is defined by the second derivative of the $N=Z$ energy per particle $E/A$ with respect to the Fermi momentum $k_\text{F}$
\begin{eqnarray}
\label{eq:comp_inf}
K_{\infty} = k^2_\text{F} \frac{d^2E/A}{dk^2_\text{F}}\Bigg|_{k_\text{F} = k_{\text{F}_0}} \,, 
\end{eqnarray}
where $k_{\text{F}_0}$ is the Fermi momentum at the saturation density $n_0$, which is defined by the stationarity condition
\begin{equation}
    \frac{dE/A}{dn}\bigg|_{n=n_0}=0\,.
\end{equation}
Equation~\eqref{eq:comp_inf} provides information about the stiffness of nuclear matter against variations in the density around the stationary point.
We introduce an effective incompressibility in a finite system of $A$ nucleons, where $\zeta$ is a representative length scale of the size of the system. The Fermi momentum $k_\text{F}$ scales as the inverse of $\zeta$, i.e.,
\begin{equation}
    k_\text{F} \sim \zeta^{-1}\,,
\end{equation}
such that Eq.~\eqref{eq:comp_inf} can be modified in a finite system as 
\begin{equation}
    K_A=\zeta^2\frac{d^2E/A}{d\zeta^2}\bigg|_{n_0}\,.
\end{equation}
If the mean square radius $\braket{r^2}$ is chosen as the typical length scale, then
\begin{equation}
\label{eq:KA_r}
    K_A=4\braket{r^2}^2\frac{d^2E/A}{d\braket{r^2}^2}\bigg|_{n_0}\,.
\end{equation}
It can be shown~\cite{Roca-Maza18a} that Eq.~\eqref{eq:KA_r} eventually leads to the expression for the incompressibility of a finite $A$-body nucleus
\begin{eqnarray}
\label{eq:KA}
K_A = \frac{m}{\hbar^2} \langle r^2 \rangle E^2_{\mathrm{GMR}}\,, 
\end{eqnarray}
where $m$ is the mass of the nucleon and $E_\mathrm{GMR}$ is the GMR energy, defined in Eq.~\eqref{eq: egmr}. The finite nucleus incompressibility $K_A$ and the nuclear matter incompressibility $K_\infty$ can be linked through a so-called leptodermous expansion, similar to the liquid-drop formula for nuclear masses,
\begin{align}
\label{eqn:KA_lepto}
K_A &= K_{\text{vol}} + K_{\mathrm{surf}} A^{-1/3} + K_{\mathrm{Coul}} Z^2 A^{-4/3}\\
\nonumber
&\quad+  K_{\mathrm{sym}} \left (\frac{N-Z}{A} \right )^2 \,,
\end{align}
where $K_{\text{vol}}, K_{\mathrm{surf}}, K_{\mathrm{Coul}}, K_{\mathrm{sym}}$ are the volume, surface, Coulomb, and symmetry contributions to the incompressibility of a finite $A$-body system, while $N$ and $Z$ are the neutron and proton numbers, respectively~\cite{Blaizot1980}. For symmetric nuclei ($N=Z$), as the ones considered in this work, the finite-nucleus $K_A$ can then be expressed as
\begin{align}
\label{eq:extrap}
    K_A = K_\text{vol} + K_\text{surf} \, A^{-1/3}\,,
\end{align}
where the Coulomb contribution has been neglected. Therefore, one can compute $K_A$ for a variety of nuclei using the above mentioned many-body methods (RPA, CC, IMSRG), and then  fit the coefficients in Eq.~(\ref{eqn:KA_lepto}), where  $K_{\mathrm{vol}}$ corresponds to $K_{\infty}$ when extrapolating to infinite $A$~\cite{Jennings1980}. This technique was extensively used to provide estimates of $K_\infty$ both from mean-field calculations~\cite{Blaizot1980} and experimental GMR data~\cite{Stone14a}. Previous works in the \textit{ab initio} community followed this approach using the SA-NCSM method~\cite{Burrows:2023ugy}, obtaining $K_{\infty} = 213(10)$\,MeV with the NNLO$_{\mathrm{opt}}$ interaction, and the PGCM employing the N$^3$LO interaction from Ref.~\cite{Huther19a}, which gave $K_{\infty} = 284(3)$\,MeV. According to Ref.~\cite{Garg18a}, this strategy can provide a useful consistency test for finite and infinite matter calculation within a given model, but large uncertainties might affect the reliability of the extrapolated values.  

\begin{figure*}[t]
    \centering
    \includegraphics[width=\linewidth]{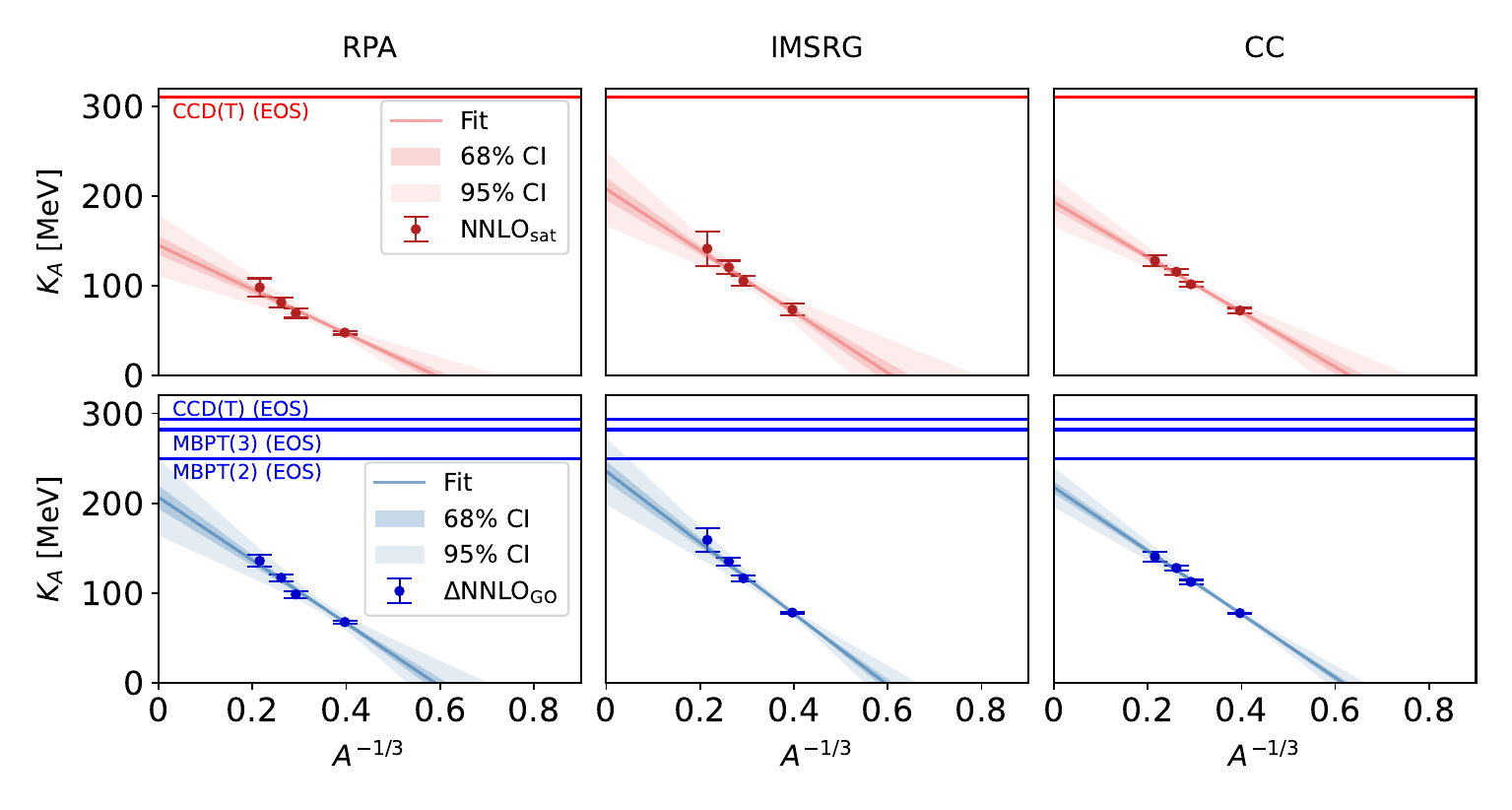}
    \caption{Linear fit from Eq.~\eqref{eq:extrap} for the RPA, IMSRG, and CC points from Fig.~\ref{fig:KA_method}. The shaded areas represent the 68\% and 95\% confidence level areas, while solid horizontal lines are numerical values for $K_\infty$ from nuclear matter calculations~\cite{Ekstrom15a,Jiang20a,Alp25a}.}
    \label{fig:KA_extrap}
\end{figure*}

In Fig.~\ref{fig:KA_method}, results for the RPA, IMSRG, and CC finite-nucleus incompressibility are displayed. Theoretical uncertainties account for the residual dependence on the model-space parameters $e_{\rm max}$ and $\hbar\omega$, and they are calculated as outlined for the values in Tab.~\ref{tab:1}. We point out the very good agreement between IMSRG and CC results for both interactions. In analogy to the average energies, \nnlosat{} yields lower $K_A$ values with respect to \deltago{}. In some instances, as in the case of $^{16}$O, the results of the two chiral interactions overlap. As noticed for moments and average energies, RPA is close to the results of correlated many-body methods in the soft interaction case, while significant deviations are seen for the harder \nnlosat{}.

The $K_A$ points from Fig.~\ref{fig:KA_method} are then used to extract the parameters from Eq.~\eqref{eq:extrap} via a linear fitting procedure. The weight of the points are chosen as $1/\sigma^2$, where $\sigma$ is the absolute uncertainty of $K_A$, stemming from its residual \Xm{e} and $\hbar\omega$ dependence. The linear fit is displayed in Fig.~\ref{fig:KA_extrap} for the three different many-body methods and for the two  interactions. The statistical uncertainty of the fit is also displayed in the form of shaded areas, indicating the 68\% and 95\% of confidence level. Results from infinite nuclear matter calculations are also displayed as horizontal lines. As already observed, the extrapolations from CC and IMSRG calculations are in good agreement, while the consistency of RPA results with the other two methods depends on the softness of the interaction, with a difference of $50-60$\,MeV for \nnlosat{} and $10-30$\,MeV for \deltago{}. The extrapolated values of $K_\infty$ are listed in Table~\ref{tab:K_infty} with the associated uncertainties.

While extrapolated IMSRG and CC results are consistent between each other, significant differences are observed with respect to infinite nuclear matter calculations based on the same interactions, see Fig.~\ref{fig:KA_extrap}.  
A sounder extrapolation would demand a larger set of nuclei, which would require the ability to access open-shell nuclei with both CC and IMSRG methods. Moreover, the Coulomb component entering Eq.~\eqref{eqn:KA_lepto} was neglected. While it is believed to have small effects in the determination of $K_\infty$~\cite{sagawa2007}, its contribution should be more deeply investigated in future developments. 
A few comments are in order also on the side of the infinite nuclear matter results. First, nuclear matter computations do not include the Coulomb interaction, which is instead included for finite nuclei. Second, a better assessment of uncertainties, particularly those stemming from the many-body truncation, would be important to draw more definitive conclusions. Finally, it is worth recalling that in nuclear matter calculations $K_{\infty}$ is obtained from the second derivative of the energy per particle with respect to the density at the saturation minimum. This extraction requires a dense sampling of the energy-density curve around the minimum to accurately approximate the second derivative. Moreover, if the interaction under consideration predicts a saturation density higher than the empirical value (e.g., for \nnlosat{} the saturation density amounts to around $0.17$\,fm$^{-3}$), this will affect the resulting $K_{\infty}$.

On the one hand, our results lie below the $250\,\MeV < K_{\infty} < 315$\,MeV range of Ref.~\cite{Stone14a}, based on a global fit of available experimental data, which, as pointed out in Sec.~\ref{sec:centroids}, can be limited in the excitation energy range they cover. On the other hand, the incompressibility values presented in this work are consistent with phenomenological ranges within their uncertainties, see, e.g.,~\cite{Roca-Maza18a}.

\begin{table}[t]
\caption{RPA, IMSRG, and CC results for the extrapolated value of $K_\infty$ from Fig.~\ref{fig:KA_extrap} for the \nnlosat{} and \deltago{} interactions.}
\begin{ruledtabular}
\begin{tabular}{r
                S[table-format=3.0(2)]
                S[table-format=3.0(2)]
                S[table-format=3.0(1)]}
$K_\infty$ [MeV] & {RPA} & {IMSRG} & {CC} \\
\hline
\nnlosat{} & 145(10) & 208(13) & 194(9) \\
\deltago{} & 206(13) & 236(11) & 218(7) \\
\end{tabular}
\end{ruledtabular}
\label{tab:K_infty}
\end{table}

\section{Conclusions and Outlook}
\label{sec:outlook}

In this work we performed a detailed comparison of the LIT-CC and IMSRG approaches applied to the calculation of the nuclear monopole response. The two frameworks yield good agreement for symmetric closed-shell nuclei using different chiral interactions, thus highlighting the consistency of the recently formulated moment-operator method~\cite{Porro25a} compared to the well-established integral transform approach. The extracted moments of the strength function allowed to calculate the finite-nucleus incompressibility, which was linked to the nuclear matter value through a leptodermous expansion. The extrapolations display sizeable differences with explicit nuclear matter calculations with the same interactions, but the values are consistent with phenomenological ranges.

For the future, we emphasize the need to account for Coulomb effects in the leptodermous expansion and to investigate asymmetric ($N\neq Z$) nuclei.
This will allow for the inclusion of more points in the extrapolation, thus increasing the fit's stability.
Moreover, the inclusion of medium-mass open-shell nuclei will be important to understand the robustness of the many-body frameworks when targeting the nuclear response in presence of strong collective correlations.

\section*{Data availability}

The data that support the findings of this article are openly available~\cite{zenodo}.

\section*{Acknowledgements}

F.B. and S.B. thank Gaute Hagen for providing access to the nuclear coupled-cluster code developed at ORNL (NUCCOR). F.B. would also like to thank Matthias Heinz for useful discussions and for providing IMSRG(3) results for moments for comparison~\cite{heinz_priv}. IMSRG results have been obtained using the \textsc{IMSRG++} code~\cite{Stroberg2024_IMSRGGit}. 
The work of F.B. is supported by the U.S. Department of Energy, Office of Science, Office of Nuclear Physics, under the FRIB
Theory Alliance award DE-SC0013617; by the U.S. Department of Energy, Office of Science, Office of Advanced Scientific Computing Research and Office of Nuclear Physics, Scientific Discovery through Advanced Computing (SciDAC) program (SciDAC-5 NUCLEI). 
The work of A.P. and A.S. is supported by the Deutsche Forschungsgemeinschaft (DFG, German Research Foundation) -- Project-ID 279384907 -- SFB 1245. 
The work of S.B. is supported by the DFG through  Project-ID 279384907 – SFB 1245, and through  the Cluster of Excellence ``Precision Physics, Fundamental Interactions, and Structure of Matter'' (PRISMA+ EXC 2118/1, Project ID 39083149).
A.T. is supported by the European Research Council (ERC) under the European Union's Horizon Europe research and innovation programme (Grant Agreement No.~101162059).

The authors gratefully acknowledge the Gauss Centre for Supercomputing e.V. (www.gauss-centre.eu) for funding this project by providing computing time on the GCS Supercomputer JUWELS at J\"ulich Supercomputing Centre (JSC). This research used resources of the Oak Ridge Leadership Computing Facility located at Oak Ridge National Laboratory, which is supported by the Office of Science of the Department of Energy under contract No. DE-AC05-00OR22725. Computer time was provided by the Innovative and Novel Computational Impact on Theory and Experiment (INCITE) program and by the supercomputer Mogon at Johannes Gutenberg Universit\"at Mainz.

\appendix
\section{Convergence of $m_1$}
\label{app:m1}

\begin{figure*}[t]
    \centering
    \includegraphics[width=\linewidth]{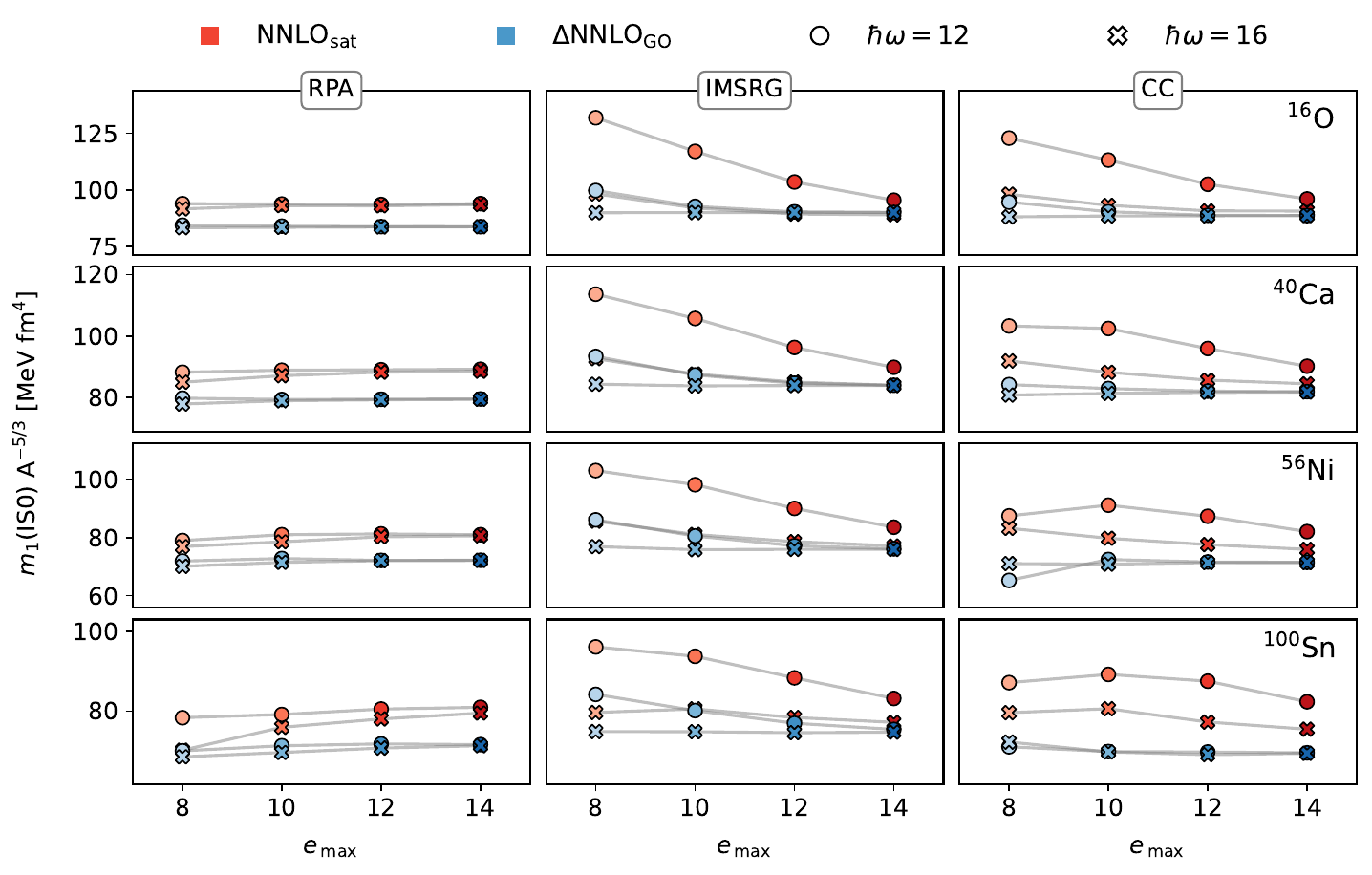}
    \caption{Values of $m_1$ for the isoscalar monopole (IS0) response for RPA (left), IMSRG (middle), and CC (right) calculations. Results are given for two different interactions (\nnlosat{}~\cite{Ekstrom15a}, \deltago{}~\cite{Jiang20a}) for $\hbar\omega=12-16$\,MeV as a function of the model space size \Xm{e}. Absolute values are rescaled by a factor $A^{5/3}$ in order to show all nuclei on the same scale.}
    \label{fig:m1}
\end{figure*}

The numerical convergence of the $m_{-1}$ and $m_0$ moments was discussed in Sec.~\ref{sec:moms_num}. In this Appendix, the $m_1$ moments for the IS0 operator from Eq.~\eqref{eq:operator} are displayed in Fig.~\ref{fig:m1}. The convergence pattern is similar to the $m_{-1}$ results from Sec.~\ref{sec:moms_num}. The model-space convergence is considerably faster for the \deltago{} interaction than for \nnlosat{} in all considered cases. When considering the most converged results, CC and IMSRG calculations are generally in good agreement. Results from RPA calculations are also not far from correlated methods, indicating that the higher-energy part of the response (since $m_1$ is an energy weighted sum), i.e., the giant monopole resonance, is less affected by the inclusion of dynamical correlations within the ground state, for both soft and hard interactions.

\bibliography{biblio}

\end{document}